\documentclass[12pt]{iopart}
\usepackage{iopams}
\expandafter\let\csname equation*\endcsname\relax
 \expandafter\let\csname endequation*\endcsname\relax
\usepackage{mathtools}
\usepackage{graphicx}
\usepackage[numbers,square,sort&compress]{natbib} 
\usepackage{enumitem}

\DeclareMathAlphabet{\mathpzc}{OT1}{pzc}{m}{it} 
\newcommand{\latin}[1]{\textit{#1}\!}
\DeclareMathOperator{\Ident}{\mathcal{I}}
\renewcommand{\Vec}[1]{\boldsymbol{\mathbf{#1}}}
\newcommand{\Oper}[1]{\mathrm{#1}}
\newcommand{\abs}[1]{\left|{#1}\right|}
\newcommand{\Z}{\mathbb{Z}}
\newcommand{\R}{\mathbb{R}}
\newcommand{\Lagrangian}{\mathcal{L}}
\newcommand{\Action}{\mathcal{S}}
\newcommand{\Kernel}{\mathcal{K}}
\newcommand{\D}[1]{\text{d}#1}
\newcommand{\PathD}[1]{\mathpzc{D}\{#1\}}
\newcommand{\Mesh}{\mathcal{C}}
\newcommand{\Weight}{\chi}
\newcommand{\Rep}{D}
\newcommand{\SecRef}[1]{section~\ref{#1}}
\newcommand{\Eqref}[1]{\eqref{#1}}
\newcommand{\Figref}[1]{figure~\ref{#1}}
\DeclareMathOperator{\Source}{\mathpzc{s}}
\DeclareMathOperator{\Target}{\mathpzc{t}}
\newcommand{\Group}[1]{\mathrm{#1}}
\DeclareMathOperator{\SpOr}{\Group{SO}} 
\begin{document}
\title[Fundamental groupoids in quantum mechanics]{Fundamental groupoids in quantum mechanics: a new approach to quantization in multiply-connected spaces}
\author{K H Neori\footnote{Corresponding author: \mailto{kneori@albany.edu}} and P Goyal}
\address{Physics Department, University at Albany, SUNY, 1400 Washington Ave, Albany, NY 12222}
\begin{abstract}
Quantization of multiply-connected spaces requires tools which take these spaces' global properties into account. Applying these tools exposes additional degrees of freedom. This was first realized in the Aharonov-Bohm effect, where this additional degree of freedom was a magnetic flux confined to a solenoid, which an electron cannot enter. Previous work using Feynman path integrals has either only dealt with specific cases, or was limited to spaces with finite fundamental groups, and therefore, was in fact inapplicable to the Aharonov-Bohm effect, as well as to interesting systems such as anyons. In this paper we start from the fundamental groupoid. This less familiar algebraic-topological object is oriented towards general paths. This makes it a more natural choice for the path integral approach than the more commonly known fundamental group, which is restricted to loops. Using this object, we provide a method that works for spaces with infinite fundamental groups. We adapt a tool used in the previous, restricted result, in order to clearly delineate physically significant degrees of freedom from gauge freedom. This allows us to explicitly build phases which take account of symmetries of the space, directly from topological considerations; we end by providing a pertinent example relating to anyons.
\end{abstract}

\submitto{\JPA}
\maketitle
\tableofcontents
\section{Introduction}
Multiply-connected spaces, that is, spaces in which not all loops can be smoothly reduced to a point, pose specific problems for quantum mechanics. This is in contrast with classical mechanics, in which global properties of a space are not relevant, since forces and fields all act locally and independently. In quantum mechanics, however, physical predictions depend on the wave-function, which spans the entirety of the space, so that such global properties become inescapable. The consequence is new degrees of freedom, whose investigation requires some subtlety.

\begin{figure}[ht!]
\centering
\includegraphics{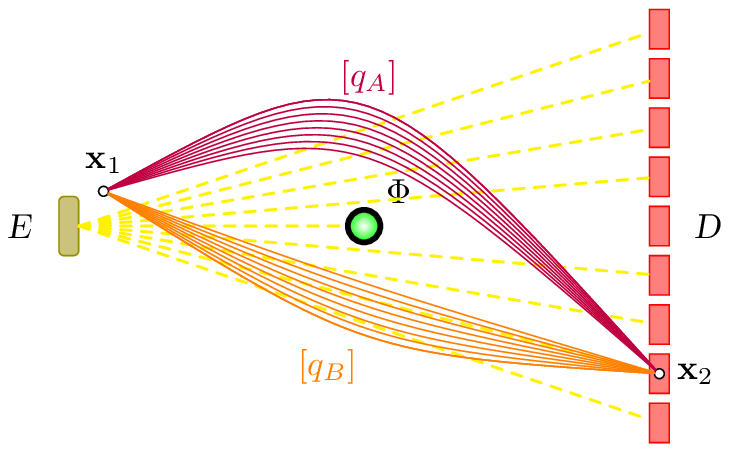}
\caption{Aharonov-Bohm apparatus. Electrons emanate from the emitter~\textsl{E} to the detectors~\textsl{D}, while being excluded from a solenoid containing a flux~$\Phi$. Two distinct families of paths between~$\Vec{x}_1$ and~$\Vec{x}_2$, which cannot be deformed into each other,  are illustrated.}\label{fig:abefeyn}
\end{figure}

The first derived physical consequence of such degrees of freedom was the~Aharonov-Bohm effect~\cite{AharonovBohm1959}. Unlike its lesser-known predecessor~\citet{EhrenbergSiday1949}, covering similar ground, it cited topological concerns explicitly. Its discoverers showed that a solenoid that is isolated from an electron can nevertheless influence it if it can move all the way around the solenoid, which is what creates a multiply-connected space. They were exploring the importance of electromagnetic potentials, and found the additional degree of freedom in the form of the flux inside the solenoid~(see~\Figref{fig:abefeyn}). 

\citeauthor{Schulman1967}~\cite{Schulman1967,Schulman1968,Schulman1971} established the method for dealing with multiply-connected spaces using the path integral approach, and~\citet{LaidlawDeWitt1971} showed how to apply Schulman's method to an abstract family of spaces. However, their approach requires them to spend considerable effort on just arriving at the algebraic structure underlying the degrees of freedom. Their result is moreover limited to spaces with a finite fundamental group. In particular, it would not apply to the Aharonov-Bohm problem, nor to anyons, which are identical particles with fractional statistics (see~\citet{Lerda1992} for an introduction. We will apply the results of this paper to this problem in~\citet{NeoriGoyal2015b}).
A final shortcoming is that it does not allow an adaptation of the topological phase to symmetries of the space.

Most papers which cite~\citet{LaidlawDeWitt1971} ignore the limitation to finite groups; see~\cite{Dowker1972,DowkerCritchley1977,CasatiGuarneri1979,GerrySingh1979,Horvathy1980a,Horvathy1980b,Horvathy1980c,Isham1983,Wu1984,Dowker1985,SudarshanImboShahImbo1988,Morandi1992}, with \citet{HorvathyMorandiSudarshan1989} an unusual exception. Nevertheless, in some cases they explicitly do apply it to the Aharonov-Bohm effect or to anyons, as in~\cite{DowkerCritchley1977,CasatiGuarneri1979,Isham1983,Wu1984,Morandi1992}.
Additionally, all of~\cite{Dowker1972,DowkerCritchley1977,CasatiGuarneri1979,GerrySingh1979,Horvathy1980a,Horvathy1980b,Horvathy1980c,Isham1983,Wu1984,Dowker1985,SudarshanImboShahImbo1988,Morandi1992,HorvathyMorandiSudarshan1989} fail to recognize the way in which this result would usually be incompatible with the symmetries of the space. \citet{Morandi1992} is notable for going further, and explicitly referring to the~\cite{LaidlawDeWitt1971} mesh framework, comparing it to a differential topological phase, but proceeding to completely ignore this issue.

The result we present in this paper lifts the restriction on applicable spaces to include infinite fundamental groups, and allows for a topological phase which directly incorporates some elements of the symmetry structure of the space. Furthermore, it simplifies the transition from path integrals to the algebraic structure. It does so by replacing the previous authors' top-down, black-box approach, which focused on the relation between the overall propagator and topologically limited partial propagators, with a bottom-up investigation starting from phases for individual paths. In the process we get to the algebraic structure, which is representations of the fundamental groupoid, very early, which simplifies the rest of the analysis. The groupoid also allows for a more natural calculation of topological phases for paths, rather than just for loops. We will also show explicitly how the framework in~\cite{LaidlawDeWitt1971} interferes with the incorporation of symmetries. Our ability to analyse spaces with infinite fundamental groups while incorporating some symmetries will be essential to our investigation in~\cite{NeoriGoyal2015b}.

The structure of the paper will be as follows:~in~\SecRef{sec:homotopy}, we will provide a short overview of homotopy, and particularly, the difference between fundamental group and groupoid; in~\SecRef{sec:exldw} we will explain the arguments in~\citet{Schulman1968} and~\citet{LaidlawDeWitt1971}, on which we wish to improve, and in particular, we will point out the problems that arise when one attempts to reconcile their framework with symmetry considerations; in~\SecRef{sec:groupoidproof}, which contains our main result, we will provide this improvement; finally, we will conclude and point to future work in~\SecRef{sec:concl}. The Appendix will show how little needs to be assumed for groupoids to have an interesting structure.
\section{Homotopy}
\label{sec:homotopy}
The most important tool used in the following analysis involves the concept of~\emph{homotopy}. Here we will give a short review of the subject, while focusing on an uncommonly used algebraic structure, the fundamental groupoid, to be defined below.

\begin{figure}[ht!]
\centering
\includegraphics{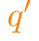}
\caption{Homotopic paths from $\Vec{a}$ to $\Vec{b}$: $[q]=[q']$.}\label{fig:homotab}
\end{figure}

We start with a space~$X$ which has some topology which we will keep implicit. Let us single out two points in that space,~$\Vec{a}$ and~$\Vec{b}$, which may be identical. A~\emph{path from}~$\Vec{a}$ \emph{to}~$\Vec{b}$ is a smooth function from the unit interval to the space:
\begin{equation}
q:[0,1]\to X
\end{equation}
such that~$q(0)=\Vec{a}$ and~$q(1)=\Vec{b}$. We then write:
\begin{equation}
\Source(q)\triangleq q(0),\quad\Target(q)\triangleq q(1),\label{eq:srctgtpathdef}
\end{equation}
the former being the path's~\emph{source}, \emph{from} which it is coming, the latter its~\emph{target}, \emph{to} which it is going. We say that two paths,~$q$ and~$q'$, with the same source and target are~\emph{homotopic} to each other if they can be smoothly deformed one into the other, while retaining the same end-points. Mathematically speaking, this means that there is a smooth function:
\begin{equation}
F:[0,1]\times[0,1]\to X
\end{equation}
such that~$F(t,0)\equiv\Vec{a}$,~$F(t,1)\equiv\Vec{b}$,~$F(0,t)\equiv q(t)$, and~$F(1,t)\equiv q'(t)$. 
This is an equivalence relation, and we can talk about the~\emph{homotopy classes} of paths between~$a$ and~$b$; that of the path~$q$ is~$[q]$, so this property can be written as:
\begin{equation}
[q]=[q']
 \end{equation}
that is, the homotopy classes of~$q$ and~$q'$ are equal~(see~\Figref{fig:homotab}).

\begin{figure}[ht!]
\centering
\includegraphics{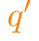}
\caption{Non-homotopic paths from $\Vec{a}$ to $\Vec{b}$: $[q]\neq[q']$.}\label{fig:nhomot}
\end{figure}

When the two paths are not homotopic, we instead write:
\begin{equation}
[q]\neq[q']\text{,}
 \end{equation}
that is, the homotopy classes of the two paths are different~(see \Figref{fig:nhomot}).

For for a homotopy class~$[q]$, we extend the definitions for~$\Source$ and~$\Target$ in~\Eqref{eq:srctgtpathdef} to:
\begin{equation}
\Source([q])\triangleq\Source(q),\quad\Target([q])\triangleq\Target(q);
\end{equation}
these are well-defined because paths in the same homotopy class share their source and target.

For convenience, if~$\Vec{x}$ is any point in X, then~$[\Vec{x}]$ is the homotopy class of the constant path at~$\Vec{x}$.

\begin{figure}[ht!]
\centering
\includegraphics{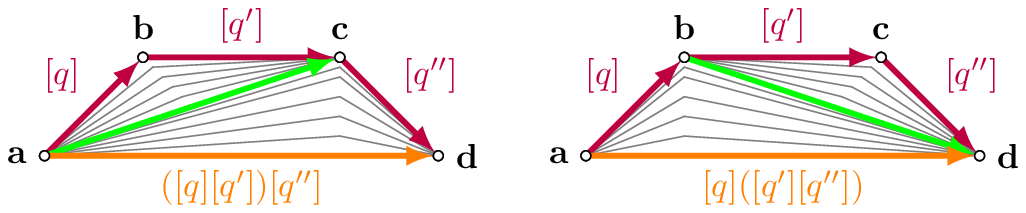}
\caption{Concatenation. Illustrating associativity:~$([q][q'])[q'']=[q]([q'][q''])$.}\label{fig:homotconcat}
\end{figure}

Given two paths,~$q$ and~$p$, such that~$\Target(q)=\Source(p)$, their \emph{concatenation}~$qp$ is well-defined as:
\begin{equation}
(qp)(t)=\begin{cases}q(2t),&0\le t\le \frac{1}{2}\\
p(2t-1),&\frac{1}{2}<t\le1
\end{cases}
\end{equation}
with~$[qp]=[q][p]$; this operation can be shown to be associative~(see~\Figref{fig:homotconcat}).
The homotopy classes with concatenation of a space~$X$ form an algebraic structure, called its~\emph{fundamental groupoid}, and written as~$\Pi(X,X)$. Let us go on a short detour to explain this structure, as it is not commonly known.

The shortest definition for a groupoid is that it is a category where every morphism is invertible. In our case, the objects of the category would be the points of the space~$X$, and the morphisms between two points are the homotopy classes of paths between them.

In more group-theoretic terms, a groupoid is a collection~$\{\alpha\}$ on which a partial operation, usually designated by a product, is defined, as well as a source function~$\Source(\bullet)$ and a target function~$\Target(\bullet)$ projecting to another space~$X$, such that:
\begin{enumerate}
\item $\alpha\beta$ is defined whenever~$\Source(\beta)=\Target(\alpha)$.
\item The operation is \emph{associative}: whenever~$(\alpha\beta)\gamma$ is defined, then it is equal to~$\alpha(\beta\gamma)$.
\item For each~$\alpha$ there is a \emph{reverse}~$\alpha^{-1}$, such that~$\alpha\alpha^{-1}=\Ident_{\Source(\alpha)}$ and~$\alpha^{-1}\alpha=\Ident_{\Target(\alpha)}$, where~$\Ident_{\Vec{x}}$ is a groupoid element called~\emph{the identity at}~$\Vec{x}$, so that~$\Source(\Ident_{\Vec{x}})=\Target(\Ident_{\Vec{x}})=\Vec{x}$, and~$\Ident_{\Source(\beta)}\beta=\beta\Ident_{\Target(\beta)}=\beta$.

Normally we would call this the~\emph{inverse}, but inversion is also a symmetry of some very important spaces, as we shall see in~\SecRef{sec:groupoidproof}. The term \emph{reverse} has the advantage of being suitable for a discussion of paths.
\end{enumerate}
For a more thorough discussion, see~\citet{Higgins1971}. A proof of the necessity of~$\Source$ and~$\Target$ is in the Appendix.

\begin{figure}[ht!]
\centering
\includegraphics{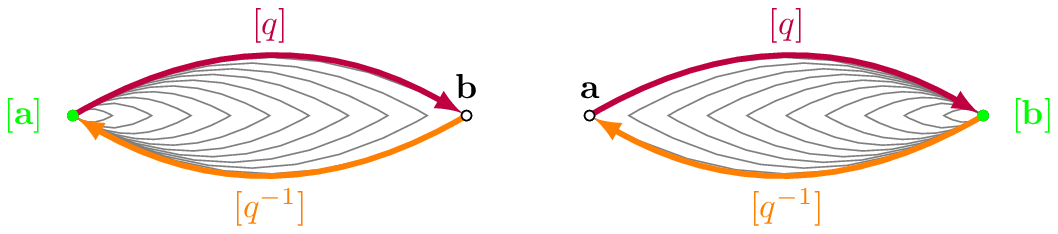}
\caption{Reverse. We see that~$[\Vec{a}]=[q][q^{-1}]$ and~$[\Vec{b}]=[q^{-1}][q]$.}\label{fig:homotreverse}
\end{figure}

We have already shown that the operation is appropriately defined and associative. To complete the list of requirements for a groupoid, we note that aside from concatenation, we also have \emph{reversal}: for each path~$q$ we may define a reverse,~$q^{-1}$, as follows:
\begin{equation}
(q^{-1})(t)=q(1-t)\text{,}
\end{equation}
and the two satisfy:
\begin{equation}
[qq^{-1}]=[\Source(q)]\quad;\quad[q^{-1}q]=[\Target(q)]\text{,}
\end{equation}
so we may define~$[q]^{-1}\triangleq[q^{-1}]$, in the sense that~$[\Vec{x}]$ acts as~$\Ident_{\Vec{x}}$~(see~\Figref{fig:homotreverse}).

\begin{figure}[ht!]
\centering
\includegraphics{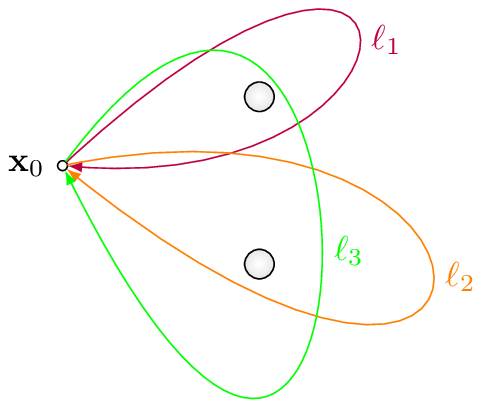}
\caption{Loops through~$\Vec{x}_0$. $[\ell_3]=[\ell_1][\ell_2]$}\label{fig:loops}
\end{figure}

Let us now choose a point~$\Vec{x}_0$ in~$X$; the fundamental \emph{group} based at that point is:
\begin{equation}
\Pi(X,\Vec{x}_0)=\left\{[q]\middle| \Source(q)=\Target(q)=\Vec{x}_0\right\}\text{;}
\end{equation}
that is, the classes of all the paths which start and finish at that point. They will also be called loops~(see~\Figref{fig:loops}).
Generally,~for a subset~$A\subseteq X$, $\Pi(X,A)$ is the subgroupoid of all~$[q]$ where~$\Source([q]),\Target([q])\in A$. Another special case used in~\SecRef{sec:exldw} is~$\Pi(X,\Vec{a},\Vec{b})$, the collection of homotopy classes~$[q]$ with~$\Source([q])=\Vec{a}$ and~$\Target([q])=\Vec{b}$.

Fundamental groups at different points are not entirely independent, as long as there is a path connecting them. This is always the case when the space is~\emph{path-connected}, which is a postulate common to~\citet{LaidlawDeWitt1971} and their successors. Indeed, let~$\Vec{x}_1$ and~$\Vec{x}_2$ be arbitrary points in~$X$. Let~$l$ be a loop based at~$\Vec{x}_1$, that is,~$\Source(l)=\Target(l)=\Vec{x}_1$, and let~$q$ be a path such that~$\Source(q)=\Vec{x}_1$ and~$\Target(q)=\Vec{x}_2$. Then~$q^{-1 }l q$ is a loop based at~$\Vec{x}_2$, and~$[q]^{-1}[l][q]$ is an element of the fundamental group there, so~$[q]$ induces a~\emph{homomorphism} between the fundamental groups, since if~$l$ and~$k$ are loops based at~$\Vec{x}_1$, then:
\begin{equation}
[q^{-1}lkq]=[q^{-1}][l][k][q]=[q^{-1}][l][q][q^{-1}][k][q]=[q^{-1}lq][q^{-1}kq]\text{;}
\end{equation}
similarly, the reverse homotopy class~$[q]^{-1}$ induces a homomorphism in the opposite direction, and:
\begin{equation}
([q]^{-1})^{-1}([q^{-1}][l][q])[q]^{-1}=[qq^{-1}lqq^{-1}]=[l];
\end{equation}
therefore, the homotopy class~$[q]$ induces an \emph{isomorphism} between the two fundamental groups. For any two homotopy classes,~$[q]$ and~$[p]$, such that~$\Source([q])=\Source([p])=\Vec{x}_1$ and~$\Target([q])=\Target([p])=\Vec{x}_2$, the respective isomorphisms induced by them are related by:
\begin{equation}
[p^{-1} l p]=[p^{-1}q q^{-1} l q q^{-1} p]=[p^{-1}q][q^{-1}l q][p^{-1}q]^{-1}\text{,}\label{eq:inneriso}
\end{equation}
which is an inner isomorphism, that is, of the form~$\alpha\to\beta\alpha\beta^{-1}$, since~$[p^{-1}q]\in\Pi(X,\Vec{x}_2)$. We will see in~\SecRef{sec:groupoidproof} that this leads to a simplification for abelian groups and for one-dimensional representations.

Finally, we can explain what we mean by simply-connected, multiply-connected, etc; always speaking of connected, path-connected spaces, a~\emph{simply-connected} space is one whose fundamental group is trivial, that is, every closed loop in it can be contracted to a point; a~\emph{multiply-connected} space is one whose fundamental group is non-trivial. So the two-dimensional plane, or any subset of it without~``holes'' is simply-connected. A space with a hole, such as the space to which electrons are restricted in the Aharonov-Bohm setup, is multiply-connected. In fact, it is~\emph{infinitely-connected}, in the sense used in~\latin{e.g.}~\cite{Schulman1968}: its fundamental group is isomorphic to an infinite group, in this case the infinite cyclic group~$\Z$ of integers.

Note that the Feynman path integral approach is very well-suited for exploring these topological properties of spaces, as paths, rather than points, are the building blocks. Indeed, some of the first works dealing with quantization in multiply-connected spaces started from this formalism. As our proof is an improvement and generalization of one such early and influential example, we turn to it next.
\section{Quantum Mechanics and the Fundamental Group}
\label{sec:exldw}
The path integral approach to multiply-connected spaces was originally introduced in~\citet{Schulman1967}, and further disseminated through~\citet{Schulman1968}. That work developed a justification for discrete spin under the path integral formalism. The configuration space which is quantized for that purpose, that of~$\SpOr(3)$ taken as a topological group, is not simply-connected, and that is where the discussion of path integrals in multiply-connected spaces becomes necessary. The standard definition of the path-integral becomes ambiguous under these circumstances: in a singly-connected space, all paths between two given points are in the same homotopy class, so they are smoothly deformable into each other, while retaining the same endpoints, and thus their contributions also go smoothly into each other. Therefore, aside from an overall phase ambiguity, the sum over paths making up the transition amplitude is defined as:
\begin{equation}
\Kernel({\Vec{b}}, t_{\Vec{b}}\,; {\Vec{a}}, t_{\Vec{a}})=\int\PathD{\Vec{x}(t)}\exp\left(i\Action\{\Vec{x}(t)\}/\hbar\right)\label{eq:pathintdef}\text{,}
\end{equation}
where~$\Vec{x}(t)$ runs over all smooth paths between~$\Vec{a}$ and~$\Vec{b}$.

If there is more than one homotopy class, then the situation is different: paths in one class can no longer be smoothly deformed into those in another, so that neither do the phases, and we in fact have to split the integral into a sum over integrals of these components:
\begin{equation}
\Kernel({\Vec{b}}, t_{\Vec{b}}\,; {\Vec{a}}, t_{\Vec{a}})=\smashoperator[r]{\sum\limits_{[q]\in\Pi(X,\Vec{a},\Vec{b})}}\widetilde{\Weight}([\Vec{x}(t)])\widetilde{\Kernel}^{[q]}({\Vec{b}}, t_{\Vec{b}}\,; {\Vec{a}}, t_{\Vec{a}})\text{,}\label{eq:amphtpcls}
\end{equation}
with the \emph{partial amplitudes} being defined as the path integral limited to the homotopy classes:
\begin{equation}
\widetilde{\Kernel}^{[q]}({\Vec{b}}, t_{\Vec{b}}\,; {\Vec{a}}, t_{\Vec{a}})\triangleq\int\PathD{\Vec{x}(t): [\Vec{x}(t)]=[q]}\exp\left(i\Action\{\Vec{x}(t)\}/\hbar\right)\text{,}\label{eq:partialintdef}
\end{equation}
and the~\emph{weights}~$\widetilde{\Weight}([q])$ yet to be determined. Note that, unlike in~\SecRef{sec:homotopy}, the parameterizations of the paths matter, which we indicate by writing them as~$\Vec{x}(t)$. The homotopy class of the path~$[\Vec{x}(t)]$ is then that of the path reparamaterized to the unit interval:
\begin{equation}
q(t)=\Vec{x}\left((t_{\Vec{b}}-t_{\Vec{a}})t+t_{\Vec{a}}\right)\text{;}
\end{equation}
we will use this notation in~\SecRef{sec:groupoidproof}, as well.

\begin{figure}[ht!]
\centering
\includegraphics{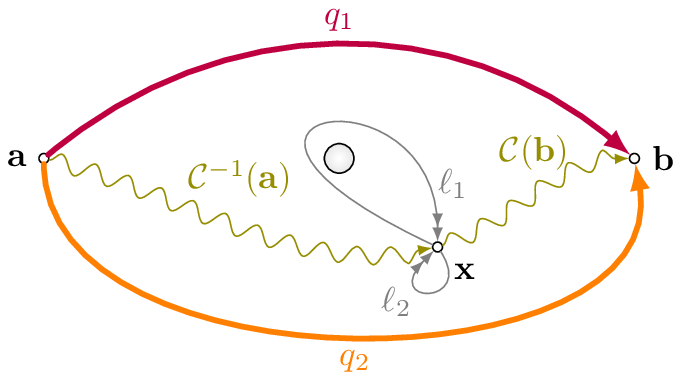}
\caption{Loops become paths through the mesh: $[q_i]=f_{\Vec{ab}}([\ell_i])=[\Mesh^{-1}(\Vec{a})][\ell_i][\Mesh(\Vec{b})]$.}\label{fig:mesh}
\end{figure}

In our adapted notation,~\Eqref{eq:amphtpcls} is the~\citet{Schulman1968} result from which~\citet{LaidlawDeWitt1971} start their analysis. However, they do not parameterize the partial amplitudes with homotopy classes, but in a roundabout way, through members of the fundamental group at some arbitrary point~$\Vec{x}_0$; hence the use of the tilde. Now, we have seen in~\SecRef{sec:homotopy} that the fundamental groups at different points for a path-connected space are related; \cite{LaidlawDeWitt1971} goes a step further, capturing \emph{all} homotopy classes, that is, members of the fundamental groupoid, by just the fundamental group at a single point,~$\Vec{x}_0$. This is done using a~\emph{mesh}, a choice of path from~$\Vec{x}_0$ to each point~$\Vec{a}$: 
\begin{align}
&\Mesh:X\to\left\{q\middle|\,q:[0,1]\to X\right\}\label{eq:meshdef}\\
&\Mesh(\Vec{a})(0)=\Vec{x}, \Mesh(\Vec{a})(1)=\Vec{a}\text{.}\notag
\end{align}
This can be done because $X$ is path-connected. Once a mesh has been chosen, a natural one-to-one correspondence between elements of the fundamental group and those of the fundamental groupoid, given the end-points, is given by:
\begin{align}
&f_{\Vec{ab}}:\Pi(X,\Vec{x})\to \Pi(X, \Vec{a}, \Vec{b})\\
&f_{\Vec{ab}}(\alpha)=[\Mesh^{-1}(\Vec{a})]\alpha[\Mesh(\Vec{b})]\text{,}
\end{align}
where $\Mesh^{-1}(\Vec{a})$ is the reverse of the path~$\Mesh(\Vec{a})$ (See~\Figref{fig:mesh}).

This works for any $\Vec{a}$ and $\Vec{b}$, allowing a consistent definition of the partial amplitudes, as well as their coefficients in the expression for the total amplitude, indexed by the fundamental group:
\begin{align}
\Kernel^{\alpha}(\Vec{b}, t_{\Vec{b}}\,; \Vec{a}, t_{\Vec{a}})&\triangleq\widetilde{\Kernel}^{f_{\Vec{ab}}(\alpha)}(\Vec{b}, t_{\Vec{b}}\,; \Vec{a}, t_{\Vec{a}})\\
\Weight(\alpha)&\triangleq\widetilde{\Weight}(f_{\Vec{ab}}(\alpha))
\intertext{providing them with an amended version of~\Eqref{eq:amphtpcls}:}
\Kernel({\Vec{b}}, t_{\Vec{b}}\,; {\Vec{a}}, t_{\Vec{a}})&=\smashoperator[r]{\sum\limits_{\alpha\in\Pi(X,\Vec{x}_0)}}\Weight(\alpha)\Kernel^{\alpha}({\Vec{b}}, t_{\Vec{b}}\,; {\Vec{a}}, t_{\Vec{a}})\text{,}\label{eq:amphtfgrp}
\end{align}
as their new starting point. For them the partial amplitudes~$\Kernel^{\alpha}$ are black boxes: they do not use the formulas~\Eqref{eq:pathintdef} or~\Eqref{eq:partialintdef}, so  a significant portion of the paper is dedicated to investigating their properties as a stepping-stone to the more relevant results constraining~$\Weight(\alpha)$. We will see in~\SecRef{sec:groupoidproof} that a direct investigation of the integrand of~\Eqref{eq:pathintdef} will save us much of this trouble.

Returning to their method, we note that $f$ respects the groupoid operation:
\begin{equation}
f_{\Vec{ab}}(\alpha)f_{\Vec{bc}}(\beta)=f_{\Vec{ac}}(\alpha\beta)\text{;}
\end{equation}
we find it helpful to see it as a parametrized local inverse to a homomorphism from the fundamental groupoid to the fundamental group at~$\Vec{x}_0$:
\begin{align}
&g:\Pi(X,X)\to\Pi(X,\Vec{x}_0)\\
&g([q])\triangleq[\Mesh(\Source([q]))][q][\Mesh^{-1}(\Target([q]))]\text{,}
\end{align}
so~$g([q][q'])=g([q])g([q'])$ and:
\begin{equation}
g(f_{\Vec{ab}}(\alpha))=\alpha\quad;\quad f_{\Source([q])\!\Target([q])}(g([q]))=[q]\text{.}
\end{equation}

The main result of their paper is a proof that the weights~$\Weight(\alpha)$ are a fundamental group representation. They then show that the choice of mesh does not change the physical predictions (it only changes the transition amplitudes by an overall phase), so that it is the choice of group representation which embodies the additional, physically meaningful topological degree of freedom.

Unfortunately, their method requires the group to be finite, limiting the realm of applicability: in particular, neither the Aharonov-Bohm effect nor anyons can be covered directly.

Furthermore, even if that were surmountable, there are fundamental limitations to their framework.  For let~$\Rep(\bullet)$ be the fundamental group representation. Then:
\begin{equation}
\widetilde{\Weight}([q])=\Rep(g([q]))\text{,}\label{eq:ldwgroupoidrep}
\end{equation}
that is, a fundamental groupoid representation has implicitly been chosen as well, through the mesh and base-point.

\begin{figure}[ht!]
\centering
\includegraphics{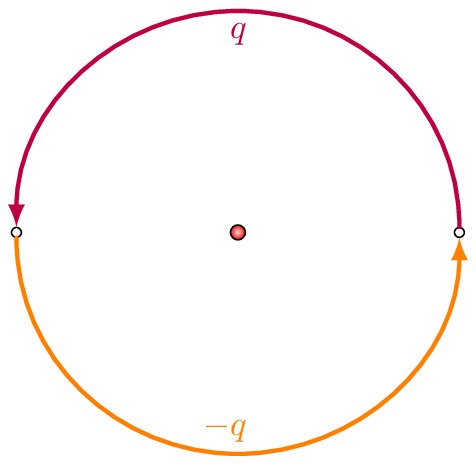}
\caption{A counter-clockwise circle is broken up into a half-circle,~$q$, and its inverse,~$-q$.}\label{fig:circletwain}
\end{figure}

We will now show that this choice does not generally allow the incorporation of symmetries of the space. We will do so for the important special case of the punctured plane,~$X=\R^2\setminus\{\Vec{0}\}$, an example we will return to in~\SecRef{sec:groupoidproof}. Again, this is under the assumption that their method could be generalized to this space, whose fundamental group at each point is isomorphic to~$\Z$ under addition, an infinite group. Let us define a ``minus'' operation, with~$-(a,b)\triangleq(-a,-b)$, where~$(a,b)$ are Cartesian coordinates. We break a circular loop counter-clockwise around the origin into two pieces,~$[q]$ and~$[-q]$, which are related by a simple inversion symmetry (not to be confused with the reversed path,~$q^{-1}$, defined in~\SecRef{sec:homotopy}):
\begin{equation}
(-q)(t)=-(q(t))\text{,}
\end{equation}
as seen in~\Figref{fig:circletwain}.

We then get:
\begin{equation}
\widetilde{\Weight}([q(-q)])=\widetilde{\Weight}([q])\widetilde{\Weight}([-q])\text{.}
\end{equation}

Now, $[q(-q)]$ generates the fundamental group at~$\Vec{x}_0$. We can then treat it as isomorphic to~$\Z$ with addition, with~$[q(-q)]$ corresponding to~$1$. Let us write:~$\widetilde{\Weight}([q(-q)])=z$. This is the generator of the image of~$\Z$ under~$\Weight(\bullet)$. In order for this representation to respect inversion symmetry, we require:
\begin{align}
\widetilde{\Weight}([-q])&=\widetilde{\Weight}([q])=w\text{,}
\intertext{so:}
z&=w^2\text{.}
\intertext{But since~$z$ generates the group, we must have~$w=z^{k}$, giving}
z&=z^{2k}\text{,}
\intertext{or}
1&=z^{2k-1}\text{.}
\end{align}
That is, this is only possible for special choices of~$z$: odd roots of unity. If we would instead like to incorporate symmetries of the space for~\emph{any} value of~$z$, we need to explore more degrees of freedom for the groupoid representation, other than re-ordering the the group representations through a different choice of mesh.

The next section will feature our main results. We will start out by proving that the phases are a groupoid representation directly, in a way that is not limited to spaces with finite fundamental groups. We will then explore what the additional degrees of freedom are, beyond the choice of a fundamental group at a point. While they create physically indistinguishable solutions, they do allow the incorporation of more of the symmetries of the space at hand. The main simplifying tool in our treatment is that its starts from paths and their phases, rather than with partial propagators. It is bottom-up rather than top-down, and does not treat propagators as black boxes.
\section{A Direct Path to the Fundamental Groupoid}
\label{sec:groupoidproof}
The aim of this section is to improve upon the work in~\citet{LaidlawDeWitt1971} in a way which uses the fundamental~\emph{groupoid}, bearing in mind its most general representations, and which is not limited to finite fundamental groups. It will be seen that the result in~\cite{LaidlawDeWitt1971} is a special case of ours, and that what they missed was a generalized gauge invariance. We will also provide an example of a groupoid representation which incorporates the rotational symmetry of the punctured plane,~$\R^2\setminus\{\Vec{0}\}$.

We start by exploring the origins of the topological degrees of freedom. Where~\citet{Schulman1967} and its successors start from the top, that is, from the propagator and the quantum formalism, we instead begin from the bottom, with Hamilton's Principle in classical physics: the least action principle, or, more accurately, the extremal action principle. 

Given an initial time~$t_1$ and point~$\Vec{x}_1$, and a final time~$t_2$ and point~$\Vec{x}_2$, we wish to find the path~$\Vec{x}_c(t)$ which is extremal (that is, a local minimum, maximum, possibly saddle) for the action functional:
\begin{equation}
\Action\{\Vec{x}(t)\}=\int\limits_{t_1}^{t_2}\Lagrangian(\Vec{x}(t),\dot{\Vec{x}}(t),t)\D{t}\text{.}
\end{equation}
We use $\{\bullet\}$ to denote a functional, rather than the usual~$[\bullet]$, since we have already reserved the latter for homotopy classes. The requirement is more formally stated as:
\begin{equation}
\delta\Action\{\Vec{x}_c(t)\}=0\text{,}
\end{equation}
with the variation keeping~$t_1,t_2,\Vec{x}_1,\Vec{x}_2$ fixed. Therefore, the same path would be extremal, even if a value depending only on these constants was added:
\begin{align}
\delta\Action\{\Vec{x}(t)\}&=\delta\left(\Action\{\Vec{x}(t)\}+\varsigma(\Vec{x}_2,t_2\,;\Vec{x}_1,t_1)\right)\text{.}
\intertext{We can go even further: extrema are only modified if the action functional is changed in such a way that its value for two paths which can be smoothly deformed into each other is altered; therefore, we can have this parameter also depend on the homotopy class of~$\Vec{x}(t)$, and write}
\delta\Action\{\Vec{x}(t)\}&=\delta\left(\Action\{\Vec{x}(t)\}+\varsigma_{[\Vec{x}(t)]}(\Vec{x}_2,t_2\,;\Vec{x}_1,t_1)\right)\text{,}
\end{align}
and this is independent of the specific Lagrangian, so long as it has the same singularity behavior (if applicable). In fact, we can write
\begin{equation}
\varsigma_{[\Vec{x}(t)]}(\Vec{x}_2,t_2\,;\Vec{x}_1,t_1)=\varsigma_{[\Vec{x}(t)]}(t_2\,;t_1)\text{,}
\end{equation}
as the end-points can be derived through~$\Source([\Vec{x}(t)])=\Vec{x}_1$ and~$\Target([\Vec{x}(t)])=\Vec{x}_2$.

Finally, let us address time dependence. We cannot rule it out entirely. For example, in the case of the punctured plane, the topological degree of freedom is equivalent to a flux through a solenoid at the origin, and that can be slowly changed, leading from one representation of the fundamental group at a point to another. Nevertheless, that does not seem to fundamentally change the form of the behavior of the system. There is wide agreement on this point in the literature, with~\cite{Dowker1972,DowkerCritchley1977,CasatiGuarneri1979,GerrySingh1979,Horvathy1980a,Horvathy1980b,Horvathy1980c,Isham1983,Wu1984,Dowker1985,SudarshanImboShahImbo1988,Morandi1992} all making a similar assumption, either implicitly, or in the case of~\citet{Dowker1972},~\citet{GerrySingh1979} and~\citet{CasatiGuarneri1979}, explicitly. We therefore choose to focus on investigating a parameter which only depends on the homotopy class of the path:
\begin{equation}
\varsigma=\varsigma([\Vec{x}(t)])\text{;}\label{eq:topparamtimeind}
\end{equation}

Now we may calculate the appropriate Feynman amplitude for the path:
\begin{equation}
\exp\left(\frac{i}{\hbar}\left(\Action\{\Vec{x}(t)\}+\varsigma([\Vec{x}(t)])\right)\right)=\exp\left(\frac{i}{\hbar}\Action\{\Vec{x}(t)\}\right)\cdot\Weight([\Vec{x}])\text{,}\label{eq:ampwithweight}
\end{equation}
where, to remain consistent with our discussion in~\SecRef{sec:exldw},~$\Weight([\Vec{x}(t)])$ is called the~\emph{weight} corresponding to the homotopy class~$[\Vec{x}(t)]$, although we remove the ``scare tilde''. In fact, since this total amplitude must obey the Feynman product rule, and since we already know that the first factor does, as it is simply that corresponding to the original action, then these weights also obey the Feynman rules. 

We are no longer interested in the time-dependence of the paths, so we can revert to the lowercase~$[q]$ notation introduced in~\SecRef{sec:homotopy}. Combining concatenation, which gives~$[qp]=[q][p]$, with the product rule of the weights gives us:
\begin{equation}
\Weight([q][p])=\Weight([qp])=\Weight([q])\Weight([p])\text{;}
\end{equation}
meaning that~$\Weight(\bullet)$ is a representation of the fundamental groupoid. Working directly with the groupoid rather than the fundamental group allows us to more generally discuss the situation for arbitrary paths, instead of being forced to work strictly through the fundamental group, as in~\SecRef{sec:exldw}. As a subgroupoid of the fundamental groupoid, we have seen that fundamental groups in two points~$\Vec{x}_1$ and~$\Vec{x}_2$ are isomorphic, with isomorphisms induced by paths, and the relation between different isomorphisms being
\begin{equation}
[p^{-1} l p]=[p^{-1}q][q^{-1}l q][p^{-1}q]^{-1}\text{;}\tag{\ref{eq:inneriso}}
\end{equation}
this is an inner isomorphism, as we stated in~\SecRef{sec:exldw}. If the group is abelian, or if we limit ourselves to a one-dimensional representation~$\Weight$, all inner isomorphisms collapse into the identity. In the former case:
\begin{align}
[p^{-1}q][q^{-1}l q][p^{-1}q]^{-1}&=[p^{-1}q][p^{-1}q]^{-1}[q^{-1}l q]=[q^{-1}l q]
\intertext{while in the latter case:}
\Weight([p^{-1}q][q^{-1}l q][p^{-1}q]^{-1})&=\Weight([p^{-1}q])\Weight([q^{-1}l q])\Weight([p^{-1}q]^{-1})=\notag\\
&=\Weight([p^{-1}q]^{-1})\Weight([p^{-1}q])\Weight([q^{-1}l q])=\Weight([q^{-1}l q])\text{,}
\intertext{so all isomorphisms between~$\Vec{x}_1$ and~$\Vec{x}_2$ induced by paths are identical. Moreover:}
\Weight([q^{-1} l q])&=\Weight([q]^{-1} [l] [q])=\Weight([q]^{-1})\Weight([l])\Weight([q])=\notag\\
&=\Weight([q])\Weight([q]^{-1})\Weight([l])=\Weight([q][q]^{-1}[l])=\Weight([\Vec{x}_1][l])=\Weight([l])\text{.}
\end{align}

This means that once a fundamental group representation is chosen at one point, it has been set for all other points. However, this is not sufficient for describing the groupoid representation, since we have only set the representations for loops, while we have to assign values to~$\Weight([q])$ when~$\Source([q])\neq\Target([q])$. As we noted in~\Eqref{eq:ldwgroupoidrep}, the framework in~\citet{LaidlawDeWitt1971} provides one way of assigning these weights. We intend to generalize that framework, using the fact that we already have the groupoid representation structure, in order to catalog all possible representations, with the advantage of allowing us to choose one that implements the symmetry of the space generally. To that end, we must introduce the notion of compatibility of representations. By saying that two representations~$\Weight$ and~$\Weight'$ of the fundamental groupoid are~\emph{compatible} we mean that, for any given~$\Vec{a}$ and~$\Vec{b}$, there is phase~$\rme^{\rmi \delta}$ such that:
\begin{equation}
\Weight'([q])=\rme^{\rmi \delta(\Vec{a},\Vec{b})}\Weight([q]),\label{eq:compatscalreps}
\end{equation}
for all~$[q]$s such that~$\Source([q])=\Vec{a}$ and~$\Target([q])=\Vec{b}$; that is, $\rme^{\rmi \delta(\Vec{a},\Vec{b})}$ only depends on the end-points.

When two representations are compatible, each one induces a propagator which results in the same physical predictions as the other, as an overall phase is removed by taking the absolute value:
\begin{equation}
\abs{\sum\limits_{[q]}\Weight'([q])\Kernel^{[q]}}=\abs{\sum\limits_{[q]}\rme^{\rmi \delta}\Weight([q])\Kernel^{[q]}}=\abs{\rme^{\rmi \delta}\sum\limits_{[q]}\Weight([q])\Kernel^{[q]}}=\abs{\sum\limits_{[q]}\Weight([q])\Kernel^{[q]}}
\end{equation}

Let us now explore exactly what kind of leeway we have with a fundamental groupoid representation once we have chosen a fundamental group representation,~$\Rep(\bullet)$. As we saw earlier, only paths between distinct points~$\Vec{a}$ and~$\Vec{b}$ can have any ambiguity to them. Let us choose a homotopy class~$[p_0]$ between~$\Vec{a}$ and~$\Vec{b}$. Then any path class~$[q]$ between these two points can be related to a loop around~$\Vec{a}$ through~$[q]\mapsto[qp_0^{-1}]$. The weight of this object is set by the fundamental group representation:
\begin{equation}
\Rep([qp_0^{-1}])=\Weight([qp_0^{-1}])=\Weight([q])\Weight([p_0]^{-1})\text{,}
\end{equation}
or:
\begin{equation}
\Weight([q])=\Rep([qp_0^{-1}])\Weight([p_0])\text{;}
\end{equation}
note that this is self-consistent, as~$\Rep([p_0p_0^{-1}])=\Rep(\Ident_{\Vec{a}})=1$.

Na{\"\i}vely it would seem that, for each pair of points~$\Vec{a}$ and $\Vec{b}$, we are free to choose an arbitrary class~$[p_{\Vec{a}\to\Vec{b}}]$, followed by a value~$\Weight([p_{\Vec{a}\to\Vec{b}}])$. 

This is not the case, however. First, this would lead to over-counting: since we have already chosen a fundamental group representation, a choice of~$\Weight([p_{\Vec{a}\to\Vec{b}}])$ is equivalent, for any other class~$[p'_{\Vec{a}\to\Vec{b}}]$ between these two points, to a choice of~$\Weight([p'_{\Vec{a}\to\Vec{b}}])$:
\begin{equation}
\Weight([p'_{\Vec{a}\to\Vec{b}}])=\Rep([p'_{\Vec{a}\to\Vec{b}}][p_{\Vec{a}\to\Vec{b}}]^{-1})\Weight([p_{\Vec{a}\to\Vec{b}}])\text{.}
\end{equation}

Furthermore, the groupoid structure imposes severe limits on our freedom. If we made the following choices: 
\begin{equation}
[p_{\Vec{a}\to\Vec{b}}],\Weight([p_{\Vec{a}\to\Vec{b}}]);\quad[p_{\Vec{b}\to\Vec{c}}],\Weight([p_{\Vec{b}\to\Vec{c}}]);\quad[p_{\Vec{a}\to\Vec{c}}],\Weight([p_{\Vec{a}\to\Vec{c}}]);
\end{equation}
then they would be subject to the constraint:
\begin{equation}
\Weight([p_{\Vec{a}\to\Vec{b}}])\Weight([p_{\Vec{b}\to\Vec{c}}])=\Weight([p_{\Vec{a}\to\Vec{b}}][p_{\Vec{b}\to\Vec{c}}])=\Weight([p_{\Vec{a}\to\Vec{c}}])\Rep([p_{\Vec{a}\to\Vec{c}}]^{-1}[p_{\Vec{a}\to\Vec{b}}][p_{\Vec{b}\to\Vec{c}}])\text{,}
\end{equation}
so even with proper counting, only two of the three can be freely chosen. 

In short, we have a big set of degrees of freedom, but they are subject to over-counting and interdependencies. Fortunately,~\citet{LaidlawDeWitt1971} furnish us with a way of organizing them: by using the base-point and mesh formalism we reviewed in~\SecRef{sec:exldw}, but in a more generalized fashion. 

If we single out any one point~$\Vec{x}_0$, and create a mesh from it to any other point,~$\Mesh(\Vec{x})$, then we can write, for any homotopy class~[q] between~$\Vec{a}$ to~$\Vec{b}$:
\begin{align}
\Weight([q])&=\Weight([\Mesh(\Vec{a})]^{-1}[\Mesh(\Vec{a})][q][\Mesh^{-1}(\Vec{b})][\Mesh(\Vec{b})])=\notag\\
&=\Weight([\Mesh(\Vec{a})]^{-1})\Weight([\Mesh(\Vec{a})q\Mesh^{-1}(\Vec{b})])\Weight([\Mesh(\Vec{b})])=\notag\\
&=\Weight([\Mesh(\Vec{a})]^{-1})\Rep([\Mesh(\Vec{a})q\Mesh^{-1}(\Vec{b})])\Weight([\Mesh(\Vec{b})])\text{;}\label{eq:oid2groupandmesh}
\intertext{we retrieve~\Eqref{eq:ldwgroupoidrep} when we choose~$\Weight([\Mesh(\Vec{x})])\equiv 1$. In particular, if we pick a different base-point,~$\Vec{x}'_0$,
and any mesh~$\Mesh'(\Vec{x})$, we have:}
\Weight([\Mesh'(\Vec{x})])&=\Weight([\Mesh(\Vec{x}'_0)]^{-1})\Rep([\Mesh(\Vec{x}'_0)\Mesh'(\Vec{x})\Mesh^{-1}(\Vec{x})])\Weight([\Mesh(\Vec{x})])\text{,}
\end{align}
so once this has been chosen for one base-point and mesh, we have set it up for all base-points and meshes. A special case of this is the one discussed in~\cite{LaidlawDeWitt1971}, a change of mesh with the same base-point. In our case we would also have to change the phases of the mesh so that~$\Weight([\Mesh(\Vec{x})])\equiv1$, which leads to a physically equivalent situation. 

Conversely, if we choose a mesh and phases for each path in the mesh (as well as the reciprocal for the reversed path:~$\Weight([\Mesh(\Vec{x})]^{-1})=\Weight^{-1}([\Mesh(\Vec{x})])$), and a representation of the fundamental group at~$\Vec{x}_0$,~\Eqref{eq:oid2groupandmesh} defines a groupoid representation; let~$q$ and~$p$ be such that~$\Source(q)=\Vec{a}$,~$\Source(p)=\Target(q)=\Vec{b}$,~$\Target(p)=\Vec{c}$; then:
\begin{align}
\Weight([q])\Weight([p])&=\Weight([\Mesh(\Vec{a})]^{-1})\Rep([\Mesh(\Vec{a})q\Mesh^{-1}(\Vec{b})])\Weight([\Mesh(\Vec{b})])\Weight([\Mesh(\Vec{b})]^{-1})\Rep([\Mesh(\Vec{b})p\Mesh^{-1}(\Vec{c})])\Weight([\Mesh(\Vec{c})])=\notag\\
&=\Weight([\Mesh(\Vec{a})]^{-1})\Rep([\Mesh(\Vec{a})q\Mesh^{-1}(\Vec{b})])\Weight([\Mesh(\Vec{b})])\Weight^{-1}([\Mesh(\Vec{b})])\Rep([\Mesh(\Vec{b})p\Mesh^{-1}(\Vec{c})])\Weight([\Mesh(\Vec{c})])=\notag\\
&=\Weight([\Mesh(\Vec{a})]^{-1})\Rep([\Mesh(\Vec{a})q\Mesh^{-1}(\Vec{b})])\Rep([\Mesh(\Vec{b})p\Mesh^{-1}(\Vec{c})])\Weight([\Mesh(\Vec{c})])=\notag\\
&=\Weight([\Mesh(\Vec{a})]^{-1})\Rep([\Mesh(\Vec{a})q\Mesh^{-1}(\Vec{b})][\Mesh(\Vec{b})p\Mesh^{-1}(\Vec{c})])\Weight([\Mesh(\Vec{c})])=\notag\\
&=\Weight([\Mesh(\Vec{a})]^{-1})\Rep([\Mesh(\Vec{a})qp\Mesh^{-1}(\Vec{c})])\Weight([\Mesh(\Vec{c})])=\notag\\
&=\Weight([qp])\text{.}
\end{align}

In fact, this is a generalization of the notion of the choice of gauge in quantum mechanics to multiply-connected spaces. Had this been a simply-connected space, this would be the same as saying we had some space-dependent phase ambiguity (as then all paths between two given points would be of the same homotopy class). More generally, there are additional degrees of freedom having to do with representations of the fundamental group at any point, which correspond to Aharonov-Bohm fluxes for the special case of spaces which can be represented as~$\R^2$ with holes.

Now, we have seen that every choice of phases for the mesh provides us with a valid groupoid representation. We have also seen that once these phases are set for one choice of base-point and mesh, the phases for any other choice are also set. We therefore proceed to prove that, given the fundamental group representation, a base-point, and a mesh, the choice of phases for the mesh has no  physical effect. Let us have two sets of mesh-weights,~$\Weight(\bullet)$ and~$\Weight'(\bullet)$, and calculate:
\begin{align}
\Weight([q])&=\Weight([\Mesh(\Vec{a})]^{-1})\Rep([\Mesh(\Vec{a})q\Mesh^{-1}(\Vec{b})])\Weight([\Mesh(\Vec{b})])=\notag\\
&=\left(\Weight([\Mesh(\Vec{a})]^{-1})\Weight([\Mesh(\Vec{b})])\right)\Rep([\Mesh(\Vec{a})q\Mesh^{-1}(\Vec{b})])\label{eq:toscalcomp1}
\intertext{and:}
\Weight'([q])&=\Weight'([\Mesh'(\Vec{a})]^{-1})\Rep([\Mesh'(\Vec{a})q\Mesh'^{-1}(\Vec{b})])\Weight'([\Mesh'(\Vec{b})])=\notag\\
&=\left(\Weight'([\Mesh'(\Vec{a})]^{-1})\Weight'([\Mesh'(\Vec{b})])\right)\Rep([\Mesh'(\Vec{a})q\Mesh'^{-1}(\Vec{b})])\label{eq:toscalcomp2}
\text{.}
\end{align}

However:
\begin{align}
\Rep([\Mesh'(\Vec{a})q\Mesh'^{-1}(\Vec{b})])&=\Rep([\Mesh'(\Vec{a})\Mesh^{-1}(\Vec{a})\Mesh(\Vec{a})q\Mesh^{-1}(\Vec{b})\Mesh(\Vec{b})\Mesh'^{-1}(\Vec{b})])=\notag\\
&=\Rep([\Mesh'(\Vec{a})\Mesh^{-1}(\Vec{a})][\Mesh(\Vec{a})q\Mesh^{-1}(\Vec{b})][\Mesh(\Vec{b})\Mesh'^{-1}(\Vec{b})])=\notag\\
&=\Rep([\Mesh'(\Vec{a})\Mesh^{-1}(\Vec{a})])\Rep([\Mesh(\Vec{a})q\Mesh^{-1}(\Vec{b})])\Rep([\Mesh(\Vec{b})\Mesh'^{-1}(\Vec{b})])=\notag\\
&=\left(\Rep([\Mesh'(\Vec{a})\Mesh^{-1}(\Vec{a})])\Rep([\Mesh(\Vec{b})\Mesh'^{-1}(\Vec{b})])\right)\Rep([\Mesh(\Vec{a})q\Mesh^{-1}(\Vec{b})])\text{.}\label{eq:toscalcomp3}
\end{align}
Note that the coefficients in larger parentheses are independent of~$[q]$, so long as~$\Source([q])$ and~$\Target([q])$ remain unchanged. Therefore, there is a phase~$\rme^{\rmi \delta}$, independent of~$[q]$, such that:
\begin{equation}
\Weight'([q])=\rme^{\rmi \delta}\Weight([q])\text{,}
\end{equation}
meaning that, unless we alter the fundamental group representation, an arbitrary change in the weights for any mesh starting from any base-point will create physically indistinguishable groupoid representations. In particular, the choice of~$\Weight([q])\equiv1$ in~\citet{LaidlawDeWitt1971} is, in fact, a legitimate special choice of gauge.

Finally, let us show that we can use this additional degree of freedom in order to directly incorporate the symmetry of a space, by returning to the case of the punctured plane,~$\R^2\setminus\{\Vec{0}\}$ from the end of~\SecRef{sec:exldw}. We will make a choice which will allow for a~\emph{single} phase,~$\rme^{\rmi \varphi}$, where~$\varphi=\phi/2$ and~$\Rep(1)=\rme^{\rmi \phi}$, for~\emph{any} half-circle counter-clockwise path around the origin.

We will (ab)use complex polar coordinates to describe points in~$\R^2\setminus\{\Vec{0}\}$:
\begin{equation}
r\rme^{\rmi \theta}\triangleq(r\cos\!\theta,r\sin\!\theta)\text{,}
\end{equation}
where~$0\le\theta<2\pi$.

\begin{figure}[ht!]
\centering
\includegraphics{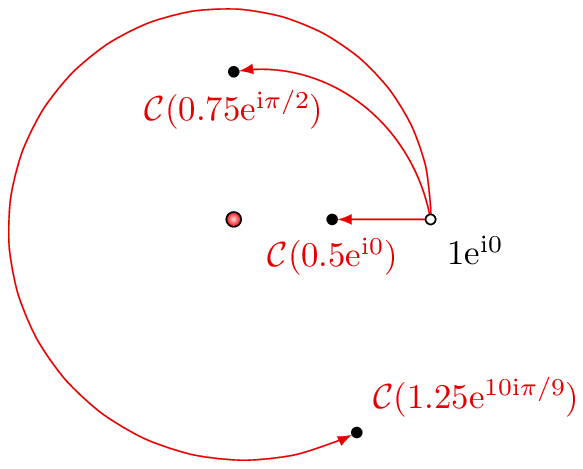}
\caption{Mesh for fundamental groupoid representation with rotational symmetry.}\label{fig:symmesh}
\end{figure}

Let~$\Vec{x}_0=(1,0)=1\rme^{\rmi 0}$, and the mesh be as follows~(see~\Figref{fig:symmesh}):
\begin{align}
\Mesh(r\rme^{\rmi \theta})(t)&=(r\rme^{\rmi \theta})^t=r^t \rme^{\rmi t\theta}\label{eq:abtopmesh}
\intertext{with the weights:}
\Weight(\Mesh(r\rme^{\rmi \theta}))&=\rme^{\rmi \phi\theta/2\pi}\text{.}\label{eq:abtopmeshweights}
\end{align}

\begin{figure}[ht!]
\centering
\includegraphics{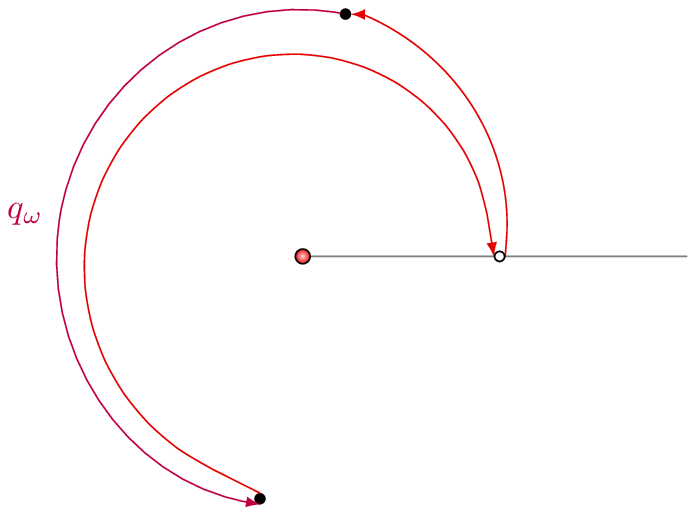}
\caption{Mesh takes a non-crossing half-circle into a contractible loop.}\label{fig:noncrossexch}
\end{figure}

\begin{figure}[ht!]
\centering
\includegraphics{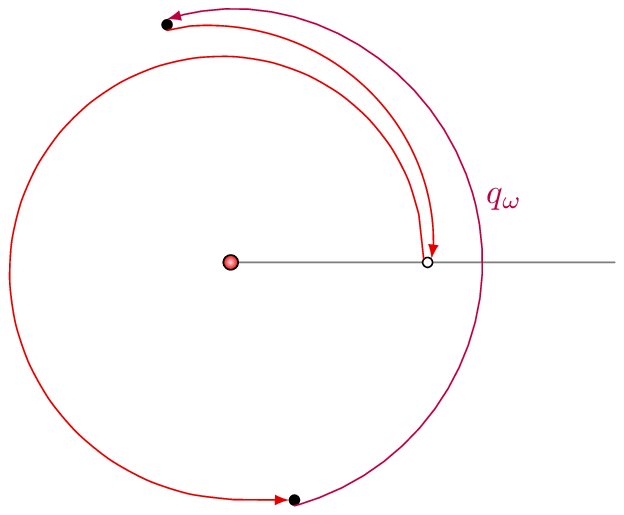}
\caption{Mesh takes a crossing half-circle into a loop with winding number~$1$.}\label{fig:crossexch}
\end{figure}

We will now show that, under this scheme, the amplitude of every counter-clockwise half-circle is equal to~$\rme^{\rmi \phi/2}$. Let~$r\rme^{\rmi \omega}$ be the starting point, and let~$q_{\omega}$ be the counter-clockwise half-circle between it and~$(-r\rme^{\rmi \omega})$. Then:
\begin{align}
\Weight(q_{\omega})&=\Weight([\Mesh(r\rme^{\rmi \omega})]^{-1})\Rep([\Mesh(r\rme^{\rmi \omega})q_{\omega}\Mesh^{-1}(-r\rme^{\rmi \omega})])\Weight([\Mesh(-r\rme^{\rmi \omega})])\text{.}
\intertext{Now, there are two possibilities: either~$0\le\omega<\pi$, in which case~$0\le\omega+\pi<2\pi$, so we can write:}
-r\rme^{\rmi \omega}&=r\rme^{\rmi (\omega+\pi)}\text{,}
\intertext{and:}
\Weight(q_{\omega})&=\rme^{-i\phi\omega/2\pi}\Rep([\Mesh(r\rme^{\rmi \omega})q_{\omega}\Mesh^{-1}(r\rme^{\rmi (\omega+\pi)})])\rme^{\rmi \phi(\omega+\pi)/2\pi}\notag\\
&=\rme^{\rmi \phi\pi/2\pi}\Rep([\Mesh(r\rme^{\rmi \omega})q_{\omega}\Mesh^{-1}(r\rme^{\rmi (\omega+\pi)})])=\rme^{\rmi \phi/2}\Rep([\Mesh(r\rme^{\rmi \omega})q_{\omega}\Mesh^{-1}(r\rme^{\rmi (\omega+\pi)})])\text{;}
\intertext{the path~$\Mesh(r\rme^{\rmi \omega})q_{\omega}\Mesh^{-1}(r\rme^{\rmi (\omega+\pi)})$ is a path which starts at~$(1,0)$, goes to~$(r,0)$, circles to~$r\rme^{\rmi \omega}$, then retraces this path back to~$(1,0)$~(see~\Figref{fig:noncrossexch}); therefore, it can be contracted to a point, and:}
\Weight(q_{\omega})&=\rme^{\rmi \phi/2}\cdot1=\rme^{\rmi \phi/2}\text{.}
\intertext{The second possibility is that~$\pi\le\omega<2\pi$, so~$0\le\omega-\pi<2\pi$, meaning we can write:}
-r\rme^{\rmi \omega}&=r\rme^{\rmi (\omega-\pi)}\text{,}
\intertext{so:}
\Weight(q_{\omega})&=\rme^{-i\phi\omega/2\pi}\Rep([\Mesh(r\rme^{\rmi \omega})q_{\omega}\Mesh^{-1}(r\rme^{\rmi (\omega-\pi)})])\rme^{\rmi \phi(\omega-\pi)/2\pi}\notag\\
&=\rme^{-i\phi\pi/2\pi}\Rep([\Mesh(r\rme^{\rmi \omega})q_{\omega}\Mesh^{-1}(r\rme^{\rmi (\omega-\pi)})])=\rme^{-i\phi/2}\Rep([\Mesh(r\rme^{\rmi \omega})q_{\omega}\Mesh^{-1}(r\rme^{\rmi (\omega-\pi)})])\text{;}
\intertext{the path~$\Mesh(r\rme^{\rmi \omega})q_{\omega}\Mesh^{-1}(r\rme^{\rmi (\omega-\pi)})$ is a path which starts at~$(1,0)$, goes to~$(r,0)$, circles clockwise to~$r\rme^{\rmi \omega}$, which causes it to \emph{cross the positive $x$-axis}, and so it is only the segment beyond which it retraces to~$(1,0)$~(see~\Figref{fig:crossexch}); therefore, it has winding number~$1$, and:}
\Weight(q_{\omega})&=\rme^{-i\phi/2}\rme^{\rmi \phi}=\rme^{\rmi \phi/2}\text{.}
\end{align}


Therefore, in both cases, which cover all counter-clockwise half-circles, the accrued phase is~$\rme^{\rmi \phi/2}$, as expected. Furthermore, any fundamental groupoid representation which agrees with this one on the fundamental group is compatible with it. Finally, it embodies rotational invariance around the origin, so it is a natural choice.
\section{Conclusion}
\label{sec:concl}
In this paper, we have found a direct derivation of the~\citet{Schulman1968} result for the Feynman path integral in multiply-connected spaces, and improved upon the treatment in~\citet{LaidlawDeWitt1971} by presenting a novel way of generating the topological degrees of freedom as representations of the fundamental groupoid. We have thereby lifted the restriction to finite fundamental groups, and have shown how this allows us to directly incorporate the symmetry of a space into the representation.

In an upcoming paper~\cite{NeoriGoyal2015b}, we will use this result to prove the strictly topological origin of anyons. The example at the end of~\SecRef{sec:groupoidproof} will be essential to this, as it provides information about paths that are not closed loops, which are necessary when one exchanges distinguishable particles.

Other future work may include multi-dimensional representations. To be more specific, if we expand compatibility in~\Eqref{eq:compatscalreps} to:
\begin{equation}
\Weight'([q])=\Oper{V}^{\dagger}(\Vec{a})\Weight([q])\Oper{V}(\Vec{b})\text{,}\notag
\end{equation}
where~$\Oper{V}(\Vec{x})$ is unitary, it promotes a similar physical equivalence for spinor or tensor propagators. Then a look at the derivations in~\Eqref{eq:toscalcomp1} to~\Eqref{eq:toscalcomp3} shows that aside from the final lines, the representation is not required to be one-dimensional or abelian, so we get compatible representations in this case, as well. This may have applications for both abelian and non-abelian gauge fields

Another possible direction would be to generalize the direct incorporation of symmetries. We further speculate that the time-dependence issue, which we avoided by restricting ourselves to~\Eqref{eq:topparamtimeind}, can best be investigated by dealing with the propagator in the space-time~$X\times\R$, which retains the connectivity properties of~$X$.
\ack
KHN would like to thank Oleg Lunin and Marco Varisco for several illuminating conversations, and Lawrence S. Schulman for helpful feedback through correspondence.
\newpage
\appendix
\section{}
\label{sec:grpdappdx}
Minimally, a groupoid is a set with an operation that is sometimes defined, such that
\begin{enumerate}
\item If~$(ab)c$ is defined then, in particular,~$ab$ is defined; futhermore,~$a(bc)$ is defined, so that~$bc$ is also defined. Finally,~$(ab)c=a(bc)$.
\item For each~$a$ there is always a~$a^{-1}$ such that~$aa^{-1}$ and~$a^{-1}a$ are defined, and:\label{it:inverse}
\begin{enumerate}[label=\alph*., ref=\alph*]
\item $(a^{-1}a)b=b$ when~$ab$ is defined\label{it:leftinverse}
\item $(aa^{-1})b=b$ when~$a^{-1}b$ is defined\label{it:leftdoubleinverse}
\item $b(aa^{-1})=b$ when~$ba$ is defined\label{it:rightinverse}
\item $b(a^{-1}a)=b$ when~$ba^{-1}$ is defined\label{it:rightdoubleinverse}
\end{enumerate}
\end{enumerate}
If we replace~\ref{it:inverse} by a weaker condition: 
\begin{enumerate}[label=\arabic*$'$.]
\setcounter{enumi}{1}
\item For each~$a$ there are~$a^{-1}_{\text{L}}$ and~$a^{-1}_{\text{R}}$ such that~$a^{-1}_{\text{L}} a$ and~$aa^{-1}_{\text{R}}$ are defined, and:
\begin{enumerate}[label=\alph*$'$.]
\item $(a^{-1}_{\text{L}} a)b=b$ when~$ab$ is defined
\item$b(aa^{-1}_{\text{R}})=b$ when~$ba$ is defined
\end{enumerate}
\end{enumerate}
we can retrieve it as follows. We note that:
\begin{equation}
a^{-1}_{\text{L}}=a^{-1}_{\text{L}} (aa^{-1}_{\text{R}})=(a^{-1}_{\text{L}} a)a^{-1}_{\text{R}}=a^{-1}_{\text{R}}\text{,}
\end{equation}
so~$a^{-1}_{\text{L}}=a^{-1}_{\text{R}}$ and we can write both as~$a^{-1}$, giving us~\ref{it:leftinverse} and~\ref{it:rightinverse}. Now, there also exists a~$(a^{-1})^{-1}$, and we find that:
\begin{equation}
a=a(a^{-1}(a^{-1})^{-1})=(a a^{-1})(a^{-1})^{-1}=(a^{-1})^{-1}\text{,}
\end{equation}
and we get~\ref{it:leftdoubleinverse} and~\ref{it:rightdoubleinverse}.~$\blacksquare$

Now, let us define:
\begin{align}
\Target(a)&=\{b: ab\text{ is defined}\}\\
\Source(b)&=\{c: c^{-1}b\text{ is defined}\}=\Target(b^{-1})
\end{align}
we can prove that~$\Target(a)=\Source(b)$ iff~$ab$ is defined: note that~$a^{-1}=\Target(a)$, since~$aa^{-1}$ is defined, and~$b\in\Source(b)$, since~$b^{-1}b$ is defined. Then:
\begin{itemize}
\item If~$\Target(a)=\Source(b)$, then~$b\in\Target(a)$, so~$ab$ is defined.
\item If~$ab$ is defined, then:
\begin{itemize}
\item If~$c\in\Source(b)$ then~$c^{-1}b$ is defined, so~$(cc^{-1})b=b$ is defined, and:
\begin{equation}
ab=a((cc^{-1})b)=(a(cc^{-1}))b=((ac)c^{-1})b\text{,}
\end{equation}
meaning~$ac$ is defined, so~$c\in\Target(a)$.
\item If~$c\in\Target(a)$ then~$ac$ is defined, so~$a(cc^{-1})=a$ is defined, and:
\begin{equation}
ab=(a(cc^{-1}))b=a((cc^{-1})b)=a(c(c^{-1}b))\text{,}
\end{equation}
meaning~$c^{-1}b$ is defined, so~$c\in\Source(b)$.
\end{itemize}
Therefore,~$\Target(a)=\Source(b)$.
\end{itemize}
And we are done.~$\blacksquare$
\addcontentsline{toc}{section}{References}
\bibliography{NeoriKlilHResearch}
\end{document}